\documentclass[aps, prl, twocolumn, superscriptaddress, showpacs]{revtex4}
\usepackage{amssymb}
\usepackage{amsmath}
\usepackage{bm}
\usepackage[dvips]{graphicx}
\usepackage{dcolumn}
\usepackage{longtable}

\begin{document}

\title{NMR study of the Mott transitions to superconductivity in the two Cs$%
_{3}$C$_{60}$ phases}
\author{Y.~Ihara}
\altaffiliation{ihara@lps.u-psud.fr}
\author{H.Alloul}
\author{P.~Wzietek}
\affiliation{Laboratoire de Physique des Solides, Universite Psris-Sud 11, CNRS UMR 8502,
91405 Orsay, France}
\author{D.~Pontiroli}
\author{M.~Mazzani}
\author{M.~Ricc{\`{o}}}
\affiliation{Dipartimento di Fisica, Universit{\`{a}} di Parma - Via G.P. Usberti 7/a,
43100 Parma, Italy}
\date{\today}

\begin{abstract}
We report an NMR and magnetometry study on the expanded intercalated
fulleride Cs$_{3}$C$_{60}$ in both its A15 and face centered cubic
structures. NMR allowed us to evidence that both exhibit a first-order Mott
transition to a superconducting (SC) state, occuring at distinct critical
pressures $p_{c}$ and temperatures $T_{c}$. Though the ground state
magnetism of the Mott phases differs, their high $T$ paramagnetic and SC
properties are found similar, and the phase diagrams versus unit volume per C%
$_{60}$ are superimposed. Thus, as expected for a strongly correlated
system, the inter-ball distance is the relvevant parameter driving the
electronic behavior and quantum transitions of these systems.
\end{abstract}

\maketitle

High transition temperature ($T_{c}$) superconductivity (SC) in the vicinity
of magnetic phases is rather commonly found nowadays, especially in
materials involving transition metal ions. There, the incidence of the spin
fluctuations on the SC state and the symmetry of its order parameter is
being highly debated and various possibilities have been considered
depending whether the electronic states involved at the Fermi level reside
on a single orbital (as in the cuprates) or multiorbital occupancy occurs
(as for Fe pnictides).

Alkali doped fullerides $A_{3}$C$_{60}$ ($A=$ alkali ion) represent a
distinct family of HTSC in a multiorbital case, as the lowest unoccupied C$%
_{60}$ molecular orbitals (named $t_{1u}$) are sixfold degenerate \cite%
{gunnarsson-RMP69}. The importance of electron correlations in $A_{n}$C$_{60}
$ compounds has been suggested first from the detection of insulating states
for even $n$, contrary to the expected metallicity for any $1\leq n\leq 5$
in an independent electron picture. On-ball localization of pairs of
electrons is favored by the energy gain due to Jahn-Teller (JT) distorsions
of the charged C$_{60}$ molecules which adds to the coulomb energy $U$. The
occurrence of this Mott Jahn-Teller insulator \cite{capone-PRB62}, in which
Hund's rule is violated, has been confirmed by the observation by NMR of a
spin gap due to singlet-triplet excitations, smaller than the optical gap,
for both crystal structures of $A_{4}$C$_{60}$ \cite{kerkoud-JPCS57} and Na$%
_{2}$C$_{60}$ \cite{brouet-PRB66}. Furthermore for $n=1$, in fcc-CsC$_{60}$,
the observation of charge segregation of singlet electron pairs on a sizable
fraction of the C$_{60}$ balls at low $T$ has been an even more direct
evidence that electron pairs are favored by JT\ distortions \cite%
{brouet-PRL82}.

For $n=3,$ JT distortions were expected to be less effective. So the $s$%
-wave SC evidenced in the fcc-$A_{3}$C$_{60}$, and the scaling of $T_{c}$
with the distance between C$_{60}$ balls has been interpreted by most
researchers as purely BCS, driven by on ball phonons, with weak incidence of
electronic correlations \cite{gunnarsson-RMP69}. The effort to increase $%
T_{c}$ by expanding the C$_{60}$ lattice has, however, led to the discovery
of various magnetic compounds, such as $($NH$_{3})_{x}$K$_{3}$C$_{60}$,
which displays a Mott transition to a SC state under pressure \cite%
{prassides-JACS121}. But one was still led to suspect that this behavior
could be attributed to a lifting of the degeneracy of the $t_{1u}$ levels by
the peculiar lattice structure required to expand the C$_{60}$ lattice \cite%
{rosseinsky-Nature364}. The most expanded fulleride Cs$_{3}$C$_{60}$ had
been found to become SC under pressure ($p$), with $T_{c}\simeq 40$ K \cite%
{palstra-SSC93}, but it has only recently been shown that this occurs in a
cubic A15 structure \cite{ganin-NM7}. A renewed interest arises then as this
phase is antiferromagnetic (AF) at ambient $p$, with $T_{N}=47$ K \cite%
{takabayashi-Science323}, and undergoes a Mott transition to a metallic
state for $p\sim 4$ kbar. It is then of great interest to find out whether
the electronic properties of A15-Cs$_{3}$C$_{60}$ exhibit any difference
with those of other fcc-$A_{3}$C$_{60}$ phases.

Ganin \textit{et al.} indicated that the low-temperature reaction process 
\cite{ganin-NM7} used to synthesize A15-Cs$_{3}$C$_{60}$ produces mixed
phases including fcc-Cs$_{3}$C$_{60}$ and body centered orthorhombic bco-Cs$%
_{4}$C$_{60}$. Using the spectroscopic capabilities of $^{133}$Cs NMR
experiments, we sorted out the signals from the two Cs$_{3}$C$_{60}$ isomers
in such mixed-phase samples.\ Taking that in advantage, we report in this
letter the first direct comparison, and demonstrate that a Mott transition
to SC occurs as well in fcc-Cs$_{3}$C$_{60}$. At ambient pressure, we
evidence the decrease of spin freezing temperature for the fcc as compared
to that of the A15 phase and associate it with the geometrical frustration
of the former lattice. The occurrence of a Mott transiton is shown to be
independent of the crystal structure and of the specific C$_{60}$ ball
orientational (merohedral) disorder present in fcc-Cs$_{3}$C$_{60}$. This
confers then a very important place to these phases in helping to reach an
understanding of SC in the vicinity of magnetic phases.

%%%%%%%%%%%%%%%%%%%%%%%%%%%%%%%  FIG1   %%%%%%%%%%%%%%%%%%%%%%%%%%%%%%%%%%%
\begin{figure}[tbp]
\begin{center}
\includegraphics[width=7.5cm]{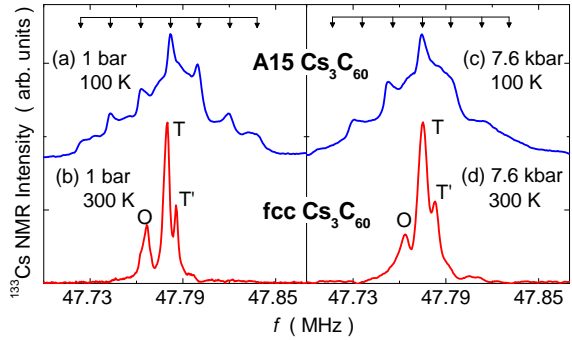}
\end{center}
\caption{a) The A15 $^{133}$Cs NMR spectrum taken at ambient pressure
displays a seven-peak powder NMR spectrum (arrows), with a quadrupole
splitting of $\sim 39$ kHz at $100$ K, and a slight asymmetric spectral
shape originating from a Knight shift anisotropy. b) $^{133}$Cs NMR spectra
for the fcc phase displaying the octahedral O and tetrahedral sites peaks, T
and T'. c) and d) The spectra are found nearly unmodified at $7.6$ kbar when
both phases are metallic (see text). }
\label{fig1}
\end{figure}
%%%%%%%%%%%%%%%%%%%%%%%%%%%%%%%  FIG1   %%%%%%%%%%%%%%%%%%%%%%%%%%%%%%%%%%%

$^{133}$Cs \textit{NMR spectra of the two phases.}--- We could synthesize
mixed-phase samples and selected three of them with significantly differing
phase contents, labelled A1 and A2 for A15 rich and F1 for the fcc rich.
Their compositions are A1\ (58.4, 12, 29.5), A2\ (41.7, 12, 46.5) , F1( 34,
55, 11), where the \% contents in formula units are given respectively for
the A15, fcc and bco phases. 
As $^{133}$Cs has a nuclear spin $I=7/2$, its NMR spectrum is
sensitive to the local site symmetry through the coupling of the nuclear
quadrupole moment with the electric field gradient (EFG) induced by the
local charge distribution. So in the A15 phase the single NMR site displays
a quadrupole-split seven-line spectrum (Fig.~\ref{fig1}a), as the unique Cs
site displays a non-cubic local symmetry \cite{jeglic-PRB80}. In the fcc
phase, the unit cell contains two alkali sites, with occupancy ratio 1:2 for
the octahedral (O) and tetrahedral (T) sites. Their local symmetry being
cubic, the EFG vanishes and each site has a narrow non-split signal. A
peculiarity evidenced by NMR \cite{walstedt-Nature362} in all formerly known
fcc-$A_{3}$C$_{60}$ is that the tetrahedral site splits into two sites (T
and T') which have been assigned to the merohedral disorder of C$_{60}$
balls \cite{matus-PRB74}. The detection in sample F1 of these three lines
(Fig.~\ref{fig1}b) establishes then the identical structure of fcc-Cs$_{3}$C$%
_{60}$. This difference in NMR spectra allowed us, as done indeed in Fig.~%
\ref{fig1}, to detect selectively the $^{133}$Cs NMR of a given phase. 
Let us point out, as will be discussed
later, that the data of Fig.~\ref{fig1}c demonstrates that these structures
are not modified under pressure. 
%%%%%%%%%%%%%%%%%%%%%%%%%%%%%%%  FIG2   %%%%%%%%%%%%%%%%%%%%%%%%%%%%%%%%%%%
\begin{figure}[tbp]
\begin{center}
\includegraphics[width=7cm]{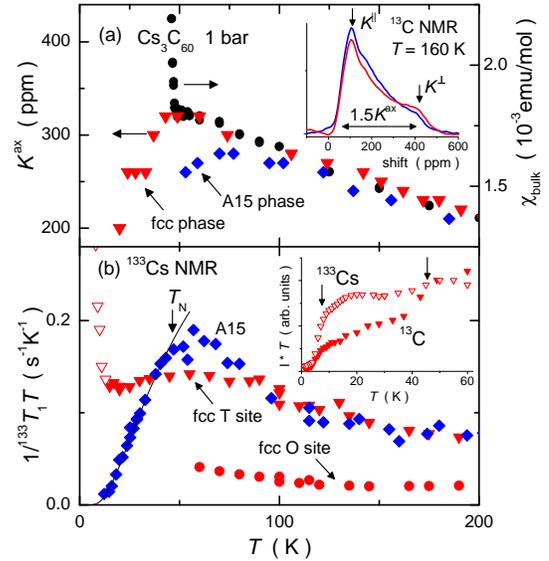}
\end{center}
\caption{(color on line) Ambient pressure data taken in samples A1 and F1
for the A15 and fcc phases. a) Similar variations above 100K\ of the
anisotropic shift contribution $K^{\mathrm{ax}}$ to the $^{13}$C spectra
(definition recalled in the inset), The scale chosen on the right for $%
\protect\chi $ (black dots, SQUID\ data on A1) emphasizes the linear
relation with $K^{\mathrm{ax}}$. b) $^{133}$Cs $(T_{1}T)^{-1}$ data for the
fcc phase T site and the A15 phase have similar high $T$ variations but
differ markedly near and below their ordered magnetic states. The spin
freezing is monitored in the inset by the variation of the $^{13}$C and $%
^{133}$Cs signal intensities taken on sample F1(see text). }
\label{fig2}
\end{figure}
%%%%%%%%%%%%%%%%%%%%%%%%%%%%%%%  FIG2   %%%%%%%%%%%%%%%%%%%%%%%%%%%%%%%%%%%

\textit{Paramagnetism at ambient pressure.}--- SQUID data on all samples do
not display any superconductivity for $p=1$ bar, and only exhibit a
paramagnetic susceptibility 
\begin{equation*}
\chi (T)=\chi _{orb}+\chi _{s}(T),
\end{equation*}%
which includes an orbital term and a $T$ dependence due to the spin
magnetism of unpaired electrons. The data were similar to those attributed
to the A15 phase \cite{takabayashi-Science323}. Comparisons between the two
phases are possible from analyses of the $^{13}$C NMR spectra, as the NMR
shift involves also orbital and spin components 
\begin{equation*}
K^{\alpha }=K_{orb}^{\alpha }+K_{s}^{\alpha }(T)=K_{orb}^{\alpha }+\
A^{\alpha }\chi _{s}(T).
\end{equation*}

Here, index $\alpha $ refers to the direction of the appplied field $B$ with
respect to local axes on the $^{13}$C site.\ Indeed, both $K_{orb}^{\alpha }$
due to the orbital magnetism of the $sp^{2}$ bonding electrons \cite%
{tycko-PRL67}, and $K_{s}^{\alpha }(T)$ associated with the spin
magnetization of electrons in the $t_{1u}\ $orbitals are anisotropic \cite%
{pennington-RMP68}. The random orientation of the balls with respect to $B$
gives a typical $^{13}$C powder NMR\ spectrum, with two singularities for $%
K^{\perp }$ and $K^{||}\ $which correspond to $B$\ directions $\perp $ and $%
||$ to the tangential plane to the C$_{60}$\ ball at the $^{13}$C\ site, as
shown in the inset of Fig.~\ref{fig2}a.\ One can notice there that, at $%
T=160 $ K, the spectra are identical for the A15\ and fcc rich samples.

Furthermore, in Fig.~\ref{fig2}a the $T$ variations of the anisotropy $K^{%
\mathrm{ax}}=2(K^{\perp }-K^{||})/3$ , which is obtained from fits of the
spectra, \textit{cannot be differenciated for the two phases above $100$ K }%
and track those of SQUID data for $\chi $ taken on sample A1. The deduced
data for $\chi _{s}(T)$ are then intrinsic and similar for both phases and
can be fitted above $100$ K with a Curie-Weiss law with an effective moment $%
p_{\mathrm{eff}}=1.52(5)\ \mu _{B}$ per C$_{60}$ and a Weiss temperature $%
\Theta _{W}=-70\pm 5$ K. Such a value for $p_{\mathrm{eff}}$ let us suggest
that the C$_{60}^{3-}$ ion is \textit{in a low spin state} in the fcc phase
as established before for the A15 phase \cite{takabayashi-Science323,
jeglic-PRB80}.

\textit{Magnetic ordering at ambient pressure.}--- For all samples the SQUID
data exhibits a sharp increase of  magnetization at $T=47$ K due to the AF
state of the fraction of A15\  phase as shown in Fig.~\ref{fig2}a. We also
detected the increased linewidth of the A15 phase $^{133}$Cs NMR signal
found in Ref.~\cite{takabayashi-Science323, jeglic-PRB80}. We could as well
find in the A15 rich sample that the magnetization on the C$_{60}$ ball
induces such a large broadening of the $^{13}$C NMR that its intensity,
detected within a small frequency window of $100$ kHz (that is $0.01$ T),
vanishes below $T_{N}=47$ K. This allowed us then in sample F1, after the
loss of the A15 fraction ($\sim 40$ \%),\  to isolate below 47K the  fcc $%
^{13}$C NMR.\ The paramagnetism of this phase,  given by  $K^{\mathrm{ax}}$, 
decreases slightly with respect to its high $T$ value, as seen in Fig.~\ref%
{fig2}a. 

%%%%%%%%%%%%%%%%%%%%%%%%%%%%%%%%%  FIG3   %%%%%%%%%%%%%%%%%%%%%%%%%%%%%
\begin{figure}[tbp]
\begin{center}
\includegraphics[width=6.5cm]{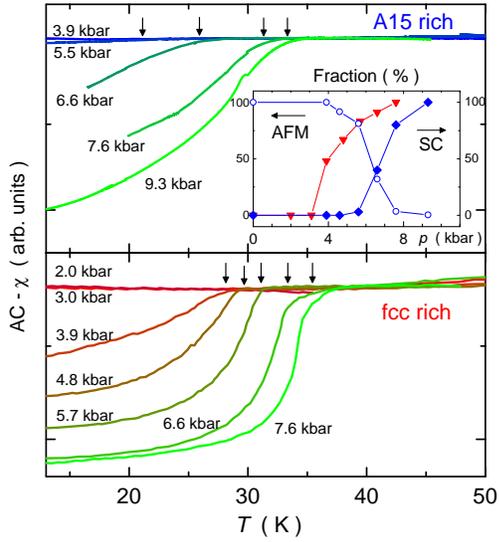}
\end{center}
\caption{(color on line) SC diamagnetism is detected at a lower $p_{c}$ for
the sample F1 than for A2 (nearly pure A15). The arrows point the values of $%
T_{c}(p)$ deduced by extrapolation of the sharp $T$ dependent diamagnetism
(slightly smaller than the onset values). The diamagnetic magnetization at
10K (extrapolated for some $p$ values) is displayed in the inset for both
samples. The magnetic volume fraction is also
shown for the A15 phase. }
\label{fig3}
\end{figure}
%%%%%%%%%%%%%%%%%%%%%%%%%%%%%%%%%  FIG3   %%%%%%%%%%%%%%%%%%%%%%%%%%%%%

Spin freezing in the fcc is only detected below $T_{f}\simeq 10$ K from the
sharp intensity drops of both the$^{13}$C and  the fcc- specific $^{133}$Cs
NMR signal, shown in the insert of Fig.~\ref{fig2}b. This latter observation
establishes that in the fcc-frozen spin state the internal field on the $%
^{133}$Cs site is much larger than that in the A15\ AF\ phase for which the
integrated intensity over a $100$ kHz window only slightly declines below $%
T_{N}$. This confirms that the transferred internal field on $^{133}$Cs is
partly compensated in the A15 AF phase due to the bipartite\ body centered
lattice symmetry \cite{jeglic-PRB80}.\ On the contrary this large internal
fields on $^{133}$Cs and the small $T_{f}\ $\ value give evidence that the
fcc phase magnetic state is influenced by the inherent frustration of the
fcc lattice.

To compare the dynamical magnetic properties, $^{133}$Cs spin lattice
relaxation $T_{1}$ data have been taken on the two phases. In the fcc, the
less shifted O site could only be resolved above $50$ K, and its $T_{1}$
scales by a factor three with that of the T site, as seen in Fig.~\ref{fig2}%
b. So both sites sense the same magnetic fluctuations through distinct
hyperfine couplings. For the T site, $(T_{1}T)^{-1}$ follows in the
paramagnetic regime a similar variation as that seen in the A15 phase.
There, below $T_{N}=47$ K \ a sharp gap in the spin excitations is detected
and the data can be fitted with $(T_{1}T)^{-1}\propto \exp (-\Delta /k_{B}T)$%
, with $\Delta \simeq 50$ K (full line on Fig.~\ref{fig2}b). On the contrary
in the fcc phase $(T_{1}T)^{-1}\ $ remains nearly constant down to $10$ K
and even displays a fast increase for the $^{133}$Cs nuclear spins which are
not submitted to a large static field. This persistence of spin fluctuations
at low $T$ is also to be linked to frustration effects. 
%%%%%%%%%%%%%%%%%%%%%%%%%%%%%%%  FIG4   %%%%%%%%%%%%%%%%%%%%%%%%%%%%%%%%%%%
\begin{figure}[tbp]
%\begin{center}
\includegraphics[width=7.5cm]{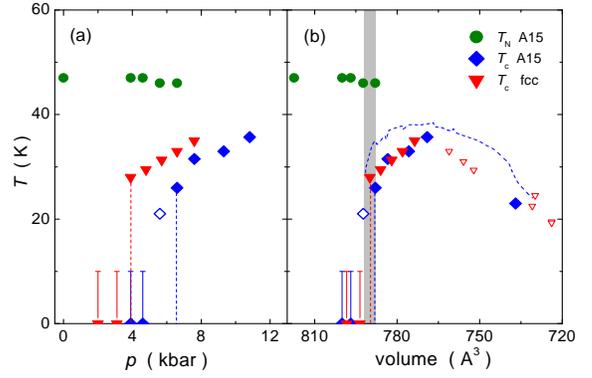} %\end{center}
\caption{(color on line) Phase diagram reporting the critical temperatures
(a) versus $p$, (b) versus $V_{C_{60}}$. The pressures $p_{c}$ of the Mott
transitions shown as dotted lines in (a) merge into the gray area in (b).
There we reported for comparison the former data on A15 phase \protect\cite%
{takabayashi-Science323} as a dotted line, and on the other fcc-$A_{3}$C$%
_{60}$ as empty downward triangles \protect\cite{maniwa-JPSJ63}. }
\label{fig4}
\end{figure}
%%%%%%%%%%%%%%%%%%%%%%%%%%%%%%%  FIG4   %%%%%%%%%%%%%%%%%%%%%%%%%%%%%%%%%%%

\textit{SC phases under pressure and phase diagram.}--- SC diamagnetism
could be probed \textit{in situ} by monitoring the shift of the NMR coil
tuning frequency ($f\simeq 28$ MHz). This allowed us to estimate the SC
volume fractions, as we kept the geometry unchanged during the pressure
sweeps. We find here that the fcc phase displays a transition to a SC state
at $3.9$ kbar (sample F1), while this occurs only above $5.5$ kbar in the
A15 rich (A2), which points out that the critical pressure $p_{c}$ for the
magnetic-SC transition is lower for the fcc phase than for the A15. This can
be seen in the insert of Fig.~\ref{fig3}, where the magnitude of the
diamagnetic signal measured at $10$ K is plotted versus $p$. It is as well
compared for the A15 phase with the magnetic volume fraction estimated from
the $^{133}$Cs NMR. The abrupt loss of magnetism above $%
7 $ kbar and onset of SC above $6$ kbar points for the \textit{first order
AF/SC phase boundaries} at $p_{c}=6.5(5)$ kbar in the A15 and $p_{c}=3.5(5)$
kbar in the fcc phase.

We can then report the $(p,T)$ phase diagrams in Fig.~\ref{fig4}a, and do
find in Fig.~\ref{fig4}b that $p_{c}$ and $T_{c}$ merge together for the two
Cs$_{3}$C$_{60}$ phases if plotted versus $V_{C_{60}}$, the unit volume per C%
$_{60}$ ball \cite{footnote}. There, we can as well compare the present $%
T_{c} $ data with former results on the other fcc-$A_{3}$C$_{60}$, and we
evidence that a similar maximum of $T_{c}$ versus $V_{C_{60}}$ applies for
the two structures.

Let us now consider the NMR data taken at high enough $p,$ for which both
phases become SC at low $T$.\ It is first clear, as seen in Fig.~\ref{fig1},
that the spectra above $T_{c}$ do not differ from those taken at one bar.
This absence of structural modification for both phases establishes then
that the evolution with $p$ only implies electronic degrees of freedom. 
%%%%%%%%%%%%%%%%%%%%%%%%%%%%%%%  FIG5   %%%%%%%%%%%%%%%%%%%%%%%%%%%%%%%%%%%
\begin{figure}[tbp]
\begin{center}
\includegraphics[width=7.5cm]{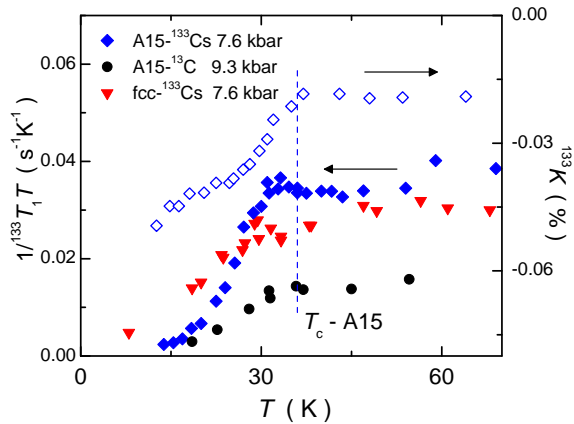}
\end{center}
\caption{$T$ variations of $(T_{1}T)^{-1}$ for $^{133}$Cs and $^{13}$C (left
scale) and of the $^{133}$Cs shift $K$ (blue empty diamonds right scale) in
the A15 phase above $p_{c}$. The magnitude of the T site $^{133}$Cs
relaxation rate in the fcc compound is found similar to that obtained in the
A15 compound (see text).}
\label{fig5}
\end{figure}
%%%%%%%%%%%%%%%%%%%%%%%%%%%%%%%  FIG5   %%%%%%%%%%%%%%%%%%%%%%%%%%%%%%%%%%%

In the metallic state, a Korringa-like $T$ independent $T_{1}T$ is seen
above $T_{c}$ for the two phases as shown in Fig.~\ref{fig5}. The constant
values are similar for $^{133}$Cs in the A15\ as for the T site of the fcc,
which displayed similar $T_{1}$ in the paramagnetic phases as well (Fig.~\ref%
{fig2}b). This points that the spin dynamics has comparable evolution with
pressure in the two phases. In the A15, the opening of the SC gap at $T_{c}$%
, as detected from the onset of decrease in the $^{133}$Cs Knight shift $%
^{133}K,$ occurs slightly above the observed decrease in $(T_{1}T)^{-1}$for
both $^{13}$C and $^{133}$Cs nuclei. This result perfectly mimics the
observations done in high applied fields in the other fcc-$A_{3}$C$_{60}$
compounds \cite{stenger-PRL74}. There, such a persistence of spin
excitations slightly below $T_{c}$ are remnants of the $s$-wave BCS like
Hebel-Slichter coherence peak \cite{coherencepeak}, which is damped in high
fields and could only be fully revealed from low field data \cite%
{gunnarsson-RMP69}.

\textit{Summary and discussion.}--- In conclusion we have shown here that
the magnetic ground states of the Cs$_{3}$C$_{60}$\ phases are quite
distinct at ambient pressure. The reduction of ordering temperature and the
persistence of low-energy spin fluctuations at low $T$ in the fcc phase can
be assigned to the frustration effects inherent to this structure and its
merohedral disorder. We further evidenced that no major crystal structure
modification occurs under pressure as shown as well from x-ray spectra in
the A15 phase \cite{takabayashi-Science323}. So the low $T$ transition from
a magnetic to a SC state, which appears to be of first order, is fully
determined by electronic parameters in both cases. Comparison of the phase
diagrams demonstrates that the crititical pressure $p_{c}$ for the
transition occurs for a similar value of the volume per C$_{60}$ ball $%
V_{C_{60}}$, which highlights the Mottness of the transition to be opposed
to a CDW/SDW transition, which should sensitively depend on lattice and
Fermi surface symmetry.

Our data evidences that fcc-$A_{3}$C$_{60}$ compounds exhibit a dome
behavior of $T_{c}(V_{C_{60}})$, identical to that found in the A15\ phase
under pressure.\ This unexpected feature in a purely BCS $s$-wave scenario
dominated by a density of state variation is then quite generic.\ Our result
gives then weight to the theoretical attempts to take correlation and JT\
effects into account in these systems \cite{capone-Science296}, which did
suggest such a behavior beforehand.\ We are presently investigating the
evolution of the spin dynamics accross the Mott transition, which should
permit more thorough comparisons with such theoretical approaches, and hope
that the present work will trigger diverse other experimental studies of the
electronic properties of the Cs$_{3}$C$_{60}$ phases across the Mott
transition.

We would like to thank V.~Brouet, M.~Fabrizio and E.~Tosatti for their
interest and for stimulating exchanges. Y.~I ackowledges financial support
from JSPS for his post-doctoral stay in Orsay.


\begin{thebibliography}{20}
\bibitem[{foo()}]{footnote}
\bibinfo{journal}{The unit cell volume per C$_{60}$, $V_{C_{60}}$ was estimated
  using the compressibilities of $5.4$ and $5.9$ ($10^{-3}$ kbar$^{-1}$) for
  the fcc and A15 phases respectively\cite{takabayashi-Science323}.}
\end{thebibliography}
\end{document}